\documentstyle[11pt,epsfig]{article}
\begin{document}
\parskip.3cm
%%% TITLE 
\noindent{\Large\bf Variations of the Core Luminosity and Solar Neutrino Fluxes}
\vskip1cm
%%% AUTHOR & AFFILIATION
\noindent{Attila Grandpierre, {\it Konkoly Observatory, Hungary}}
\vskip.3cm
%%% E-MAIL
\noindent{email: {\tt Grandp@konkoly.hu}}
\vskip.1cm
\noindent 
The aim 
of the present work is to analyze the geological and astrophysical 
data as well as presenting theoretical considerations indicating the presence 
of dynamic processes present in the solar core. The dynamic solar model 
(DSM) is suggested to take into account the presence of cyclic variations in 
the temperature of the solar core.
Comparing the results of calculations of the CO$_2$ content, albedo and solar 
evolutionary luminosity changes with the empirically determined global 
earthly temperatures, and taking into account climatic models, I determined 
the relation between the earthly temperature and solar luminosity. These 
results indicate to the observed maximum of $10^\circ$ 
change on the global terrestrial surface temperature a related 
solar luminosity change around $4-5$ \% 
on a ten million years timescale, which 
is the timescale of heat diffusion 
from the solar core to the surface. The related solar core temperature 
changes are around 1 \% only. At the same time, the cyclic 
luminosity changes of the solar core are shielded effectively by the outer 
zones since the radiation diffusion takes more than $10^5$ years to reach the 
solar surface. The measurements of the solar neutrino fluxes with 
Kamiokande 1987-1995 showed variations higher than 40 \% around the 
average, at the Super-Kamiokande the size of the apparent scatter decreased 
to 13 \%. This latter scatter, if would be related completely to stochastic 
variations of the central temperature, would indicate a smaller than 1 \% 
change. Fourier and wavelet analysis of the solar neutrino fluxes indicate 
only a marginally significant period around 200 days (Haubold, 1998). 
Helioseismic measurements are known to be very constraining. Actually, 
Castellani et al.\ (1999) remarked that the different solar models lead to 
slightly different sound speeds, and the different methods of regularization 
yield slightly different sound speeds, too. Therefore, they doubled the found 
parameter variations, and were really conservative assuming that errors add 
up linearly. This conservative error estimation gives 
$\delta u/u = 1.7$ \%, $\delta \rho/\rho = 
7$ \% at $r=0.06\times R_\odot$, and so the $\delta T/T = 9$ \%, 
since $\delta T/T \sim \delta \rho/\rho + \delta P/P.$ At
 $r=0.04\times R_\odot$,
$\delta u/u=2.2$ \%, $\delta \rho/\rho=10$ \%, 
$\delta T/T=13$ \%. At $r=0,$ $\delta u/u=3.5$ \%, therefore 
$\delta \rho/\rho=16$ \% 
and so $\delta T/T=20$ \%. 
So even with the usual, not conservative error 
estimation, roughly dividing these conservative errors by 4, with 
$\delta u/u=0.4$ \%, 
we still get an allowed range cca. 2 \% temperature change at 
$r=0.06\times R_\odot$ and higher in the more central regions.

In solar core varying cyclically on a decade timescale, the longer timescale 
nuclear reactions cannot build up equilibrium. In such a short timescale the 
variations of the local temperature regulates the proton-proton chain instead 
of the global luminosity constraint that is applicable only on evolutionary 
timescales. Therefore, the temperature dependences of the pp cycle neutrinos 
will be different from the ones determined by solar model calculations with 
the luminosity constraint: instead of the usual\vskip-3mm
$$
pp \sim T^{-1/2},\ \  {\rm Be} \sim T^8,\ \  {\rm B}\sim T^{18}.	
$$\vskip-1mm
\noindent we determined by the nuclear reaction rates formulas\vskip-3mm
$$
pp \sim T^{4.2},\ \  {\rm Be} \sim T^{-1/2},\ \  {\rm B}\sim T^{13.5},
$$ 
for $\tau < 10^2$ years.
These latter relations have high significance at estimating the uncertainties 
of the solar central temperatures without assuming the luminosity constraint.
Although the purely astrophysical solutions seem to be ruled out, this is not 
the case for a model in which astrophysical effects are included besides the 
neutrino oscillations. Therefore a combined, DSM+MSW model is 
suggested to calculate the observed solar neutrino fluxes. 
At present we have three types of neutrino detectors, and they offer us the 
data as the total rates (3 measurements), zenith angle dependences, energy 
spectra and day-night variations, all together 6 kind of data. The highest 
statistical significance is found in the total rates data. 
The evaluation of these 
6 data sets is not straightforward. For example, the combined fits to the 
rates+spectra+D/N changes give a bad fit to the total rates, indicating the 
need to include the astrophysical factors besides the MSW effect. 
The DSM suggest that the core dynamics is induced by intermittent events 
of dissipation of rotational energy in the solar core, in relation to angular 
momentum dissipation arising from the relative motion of the Sun and the 
mass center of the Solar System. Energetic estimations show the plausibility 
of the suggested mechanisms. 
The DSM may serve as a useful tool to describe the observed neutrino 
fluxes, shifting the allowed ranges of the MSW parameters into a more 
acceptable region. The role of the astrophysical factors in the solar neutrino 
problem is behind the fact why the ``smoking guns'' of neutrino oscillations 
are not found yet.

\end{document}